\documentclass[sigconf]{acmart}
\AtBeginDocument{%
  }

\setcopyright{acmlicensed}
\copyrightyear{2018}
\acmYear{2018}
\acmDOI{XXXXXXX.XXXXXXX}
\acmConference[Conference acronym 'XX]{Make sure to enter the correct
  conference title from your rights confirmation email}{June 03--05,
  2018}{Woodstock, NY}
\acmISBN{978-1-4503-XXXX-X/2018/06}

\usepackage{tcolorbox}
\tcbuselibrary{listings,skins,breakable}
\usepackage{listings}
\usepackage{tabularx}
\usepackage{booktabs}
\usepackage{multirow}
\usepackage{amsmath}
\usepackage{xcolor,color,soul}
\usepackage[table]{xcolor}
\lstdefinelanguage{yaml}{
  basicstyle=\ttfamily\footnotesize,
  sensitive=false,
  comment=[l]{\#},
  commentstyle=\color{gray},
  stringstyle=\color{black},
  moredelim=**[is][\bfseries]{**}{**}
}

\newtcblisting{PromptBox}[2][]{%
  breakable,
  enhanced,
  colback=gray!6,
  colframe=gray!60,
  colbacktitle=gray!60,
  coltitle=black,
  fonttitle=\bfseries,
  title={#2},
  boxed title style={
    sharp corners,
    boxrule=0pt,
    left=2mm, right=2mm, top=0mm, bottom=0mm
  },
  boxrule=0.6pt,
  arc=1mm,
  left=2mm, right=2mm, top=0mm, bottom=0mm,
  listing only,
  listing options={
    language=yaml,
    keepspaces=true
  },
  #1
}

\newtcblisting{JsonBox}[2][]{%
  breakable,
  enhanced,
  colback=gray!6,
  colframe=gray!60,
  colbacktitle=gray!60,
  coltitle=black,
  fonttitle=\bfseries,
  title={#2},
  boxed title style={
    sharp corners,
    boxrule=0pt,
    left=0mm, right=0mm, top=0mm, bottom=0mm
  },
  boxrule=0.6pt,
  arc=1mm,
  left=2mm, right=2mm, top=0mm, bottom=0mm,
  listing only,
  listing options={
    language=yaml,
    keepspaces=true
  },
  #1
}

\begin{document}

\title{Paper2Data: Large-Scale LLM Extraction and Metadata Structuring of Global Urban Data from Scientific Literature}


\author{Runwen You}
\affiliation{%
  \institution{Jilin University}
   \institution{Zhongguancun Academy}
      \country{Beijing, China}
      }

\author{Tong Xia*}
\affiliation{%
  \institution{Tsinghua University}
      \country{Beijing, China}
      }

\author{Jingzhi Wang, Jiankun Zhang, Tengyao Tu}
\affiliation{%
  \institution{Zhongguancun Academy}
      \country{Beijing, China}
      }

\author{Jinghua Piao}
\affiliation{%
  \institution{Tsinghua University}
      \country{Beijing, China}
      }

\author{Yi Chang}
\affiliation{%
  \institution{Jilin University}
      \country{Jilin, China}
      }

\author{Yong Li*}
\affiliation{%
  \institution{Tsinghua University}
   \institution{Zhongguancun Academy}
      \country{Beijing, China}
    \authornote{Corresponding authors: \{tongxia, liyong07\}@tsinghua.edu.cn}
      }
\renewcommand{\shortauthors}{Trovato et al.}
\newcommand{\tx}[1]{\sethlcolor{pink}\hl{[Tong: #1]}}

\begin{abstract}
Urban data support a wide range of applications across multiple disciplines.
However, at the global scale, there is no unified platform for urban data discovery.
As a result, researchers often have to manually search through websites or scientific literature to identify relevant datasets.
To address this problem, we curate an open urban data discovery portal, \textit{UrbanDataMiner}, which supports dataset-level search and filtering over more than 60{,}000 urban datasets extracted from over 15{,}000 Nature-affiliated publications. \textit{UrbanDataMiner} is enabled by \textit{Paper2Data}, a novel large-scale LLM-driven pipeline that automatically identifies dataset mentions in scientific papers and structures them using a unified urban data metadata schema.
Human-annotated evaluation demonstrates that \textit{Paper2Data} achieves high recall (approximately 90\%) in dataset identification and high field-level precision (above 80\%).
In addition, \textit{UrbanDataMiner} can retrieve over 9\% of datasets that are not easily discoverable through general-purpose search engines such as Google.
Overall, our work provides the first large-scale, literature-derived infrastructure for urban data discovery and enables more systematic and reusable data-driven research across disciplines.
Our code and data are publicly available\footnote{\url{https://github.com/Yourunwen/Paper2Data}}.

\end{abstract}



\received{20 February 2007}
\received[revised]{12 March 2009}
\received[accepted]{5 June 2009}

\maketitle

\vspace{-2mm}
\section{Introduction}

Cities concentrate population, infrastructure, and economic activities, making them the primary spatial units where environmental, social, and behavioral processes intersect.
As a result, urban data are inherently diverse and heterogeneous, spanning environmental conditions~\cite{jorgensen2013handbook}, population health~\cite{mooney2018bigdata}, transportation and mobility~\cite{torre2018bigdata}, and social and economic dynamics~\cite{foster2016bigdata}.
These data underpin a wide range of analytical tasks~\cite{townsend2015citiesofdata,creutzig2019upscaling}, forming a shared empirical foundation for cross-domain and multidisciplinary research.

Despite their central role, \textbf{urban datasets lack a unified, dataset-level discovery and access infrastructure at the global scale}. Many valuable datasets are not released as standalone research artifacts, but are instead embedded within scientific publications, supplementary materials, project webpages, or institutional repositories, contrary to the FAIR principles~\cite{wilkinson2016fair}. Consequently, identifying relevant urban datasets typically requires manually scanning large volumes of literature~\cite{chapman2020dataset}, interpreting heterogeneous narrative descriptions, and inferring data suitability on a case-by-case basis. This is an ad hoc and labor-intensive process that substantially hinders large-scale reuse and integration.

Existing mechanisms, such as Data Availability Statements and general-purpose search engines, primarily operate at the level of documents or webpages rather than datasets. While these approaches may indicate whether data exist, they do not support systematic, dataset-level indexing or comparison when datasets are described only in unstructured scientific text. As a result, a large fraction of urban datasets documented in the literature remain effectively invisible to dataset-oriented discovery and reuse, underscoring the absence of a unified global infrastructure for urban data discovery.

\begin{figure}[t]
  \includegraphics[width=\columnwidth]{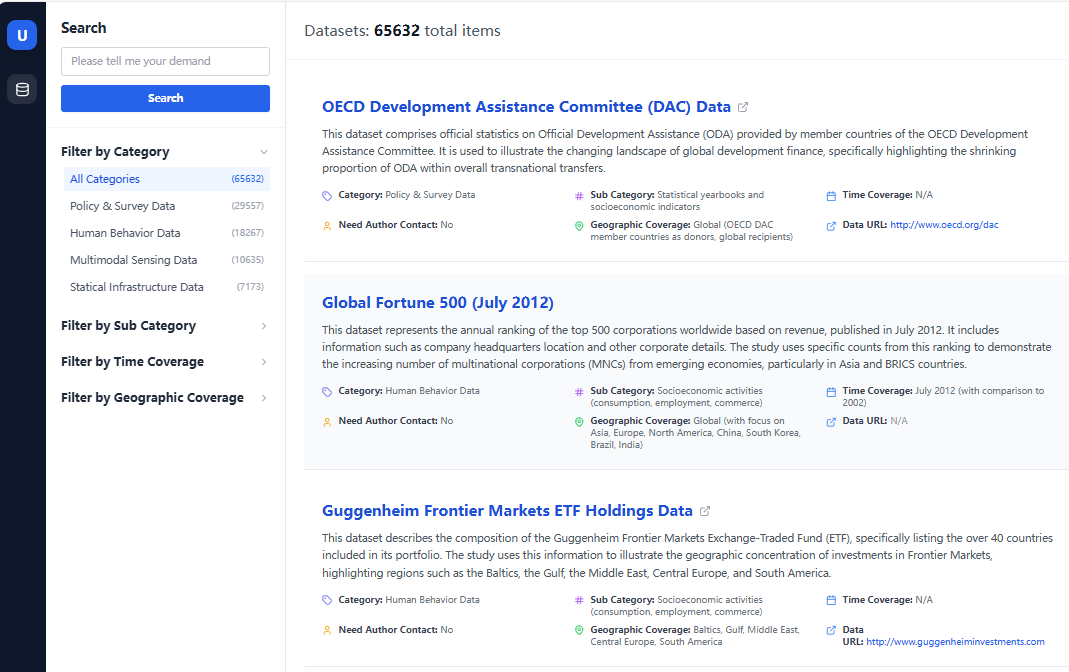}
  \caption{UrbanDataMiner portal overview. 60K+ urban datasets are provided to facilitate research use.}
  \label{fig:portal}
  \vspace{-5mm}
\end{figure}

\textbf{Recent advances in large language models (LLMs) create a new opportunity to address this long-standing gap.}
Modern LLMs exhibit strong capabilities in long-context understanding, information extraction, and semantic normalization, enabling them to identify dataset mentions embedded in narrative scientific text and to infer structured attributes such as spatial coverage, temporal scope, thematic focus, and access conditions~\cite{xu2025chatpd}. These capabilities make it feasible, for the first time, to systematically transform unstructured dataset descriptions in large-scale scientific literature into structured, machine-readable records, which bridge the gap between document-centric publication practices and dataset-centric discovery needs.

Building such a dataset-level infrastructure, however, remains challenging.
\textit{First, urban data lack a unified metadata schema and classification system.}
Datasets vary widely in modality, spatial scale, temporal coverage, and thematic focus, yet these properties are often described using inconsistent terminology across studies, making aggregation and comparison difficult.
\textit{Second, dataset descriptions are typically coarse-grained and ambiguous.}
Even after identifying a relevant paper, researchers must manually interpret narrative text to infer key dataset attributes, substantially increasing the cost of discovery.
\textit{Third, existing search paradigms remain document-centric.}
Datasets described only within scientific articles cannot be reliably indexed or retrieved using conventional dataset search tools.

To address these challenges, we propose \textbf{Paper2Data}, a literature-driven framework that systematically transforms unstructured dataset descriptions in scientific publications into structured and searchable urban data records. 
Paper2Data is explicitly designed to leverage LLMs to (i) identify dataset mentions at the article level, (ii) extract and normalize key metadata using a unified urban data schema, and (iii) model datasets as first-class objects for indexing and retrieval, rather than treating papers as the primary retrieval unit.
Based on the extracted records, we further build \textbf{UrbanDataMiner} (as shown in Figure~\ref{fig:portal}), an open urban data discovery portal that enables dataset-level search and multi-dimensional filtering.

Our main contributions in this paper are summarized as follows:
\begin{itemize}
    \item \textbf{An LLM-driven dataset identification and metadata extraction framework.}
    We design an end-to-end pipeline that automatically identifies dataset mentions in scientific literature and extracts structured metadata using a unified urban data schema, capturing key attributes required for dataset-level discovery, including spatial coverage, temporal scope, thematic category, and access conditions.

    \item \textbf{A large-scale, literature-derived urban data corpus with an open discovery interface.}
    We apply Paper2Data to approximately 15,000 Nature-affiliated publications from the past decade and construct a curated corpus of over 60,000 urban dataset records spanning diverse geographic regions, temporal ranges, and thematic categories. All records are organized using the unified metadata schema and are made publicly accessible through an open discovery portal.

    \item \textbf{Evaluation and macro-level analysis of urban data in the literature.}
    Through human-annotated evaluation, we demonstrate that Paper2Data achieves approximately 90\% recall in dataset identification and above 80\% accuracy across key metadata fields. \textit{UrbanDataMiner} can retrieve over 9\% of datasets that are not easily discoverable through general-purpose search engines. We further analyze the spatial, thematic, and temporal distributions of urban datasets in the corpus, revealing macro-level research trends and persistent data gaps across domains.
\end{itemize}

The remainder of this paper is organized as follows.
Section~\ref{sec:background} reviews related work on dataset discovery and scientific literature mining.
Section~\ref{sec:pipeline} introduces the design and implementation of the Paper2Data framework.
Section~\ref{sec:results} reports evaluation results, and Section~\ref{sec:data} presents a large-scale characterization and analysis of the extracted urban data corpus.

\section{Related Work}\label{sec:background}

\subsection{Urban Datasets and Discovery Bottleneck}
The rapid advancement of urban computing has been driven by the availability of task-specific data such as SpaceNet~\cite{van2018spacenet}, DeepGlobe~\cite{demir2018deepglobe}, and the OpenCities AI Challenge~\cite{gfdrr2020opencities}. 
These datasets establish high-quality ``gold standards'' for narrowly defined tasks, including building footprint extraction and road network mapping from high-resolution remote sensing imagery.
More recently, UrbanDataLayer~\cite{wang2024urbandatalayer} proposed a unified data pipeline and standardized data structures to support multimodal urban data engineering.
Despite these advances, existing urban data repositories remain largely task-centric and geographically constrained.
They are typically optimized for predefined benchmarks or schemas and do not support open-ended data discovery across disciplines.
In practice, a substantial fraction of urban datasets remains embedded within scientific publications rather than deposited in public repositories.
Prior studies suggest that even in mature areas such as spatiotemporal prediction, fewer than $30\%$ of papers release their underlying data~\cite{Semmelrock2025reproducibility}.
As a result, researchers must manually sift through large volumes of literature to identify relevant datasets, a process that is labor-intensive and poorly supported by keyword-based academic search engines, which struggle to bridge the semantic gap between user intent and unstructured scientific text.

\begin{figure*}[t]
  \centering
  \includegraphics[width=1.95\columnwidth]{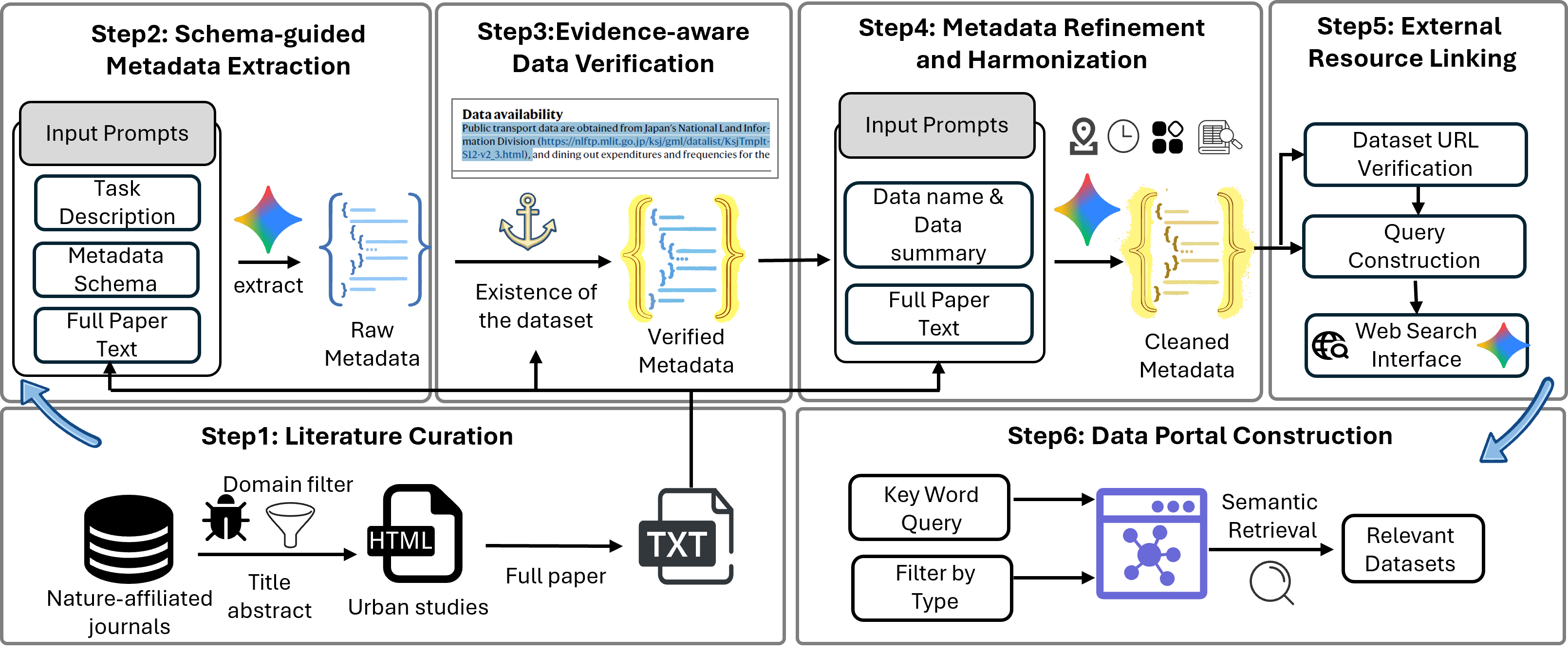}
\caption{Overview of the \textit{Paper2Data} pipeline. The system consists of six steps that transform scientific literature into a data portal where urban data can be easily retrieved.}
  \label{fig:fram}
\vspace{-2mm}
\end{figure*}

\subsection{Scientific Information Extraction and LLM-Driven Pipelines}
To address the growing discovery bottleneck, scientific information extraction (IE) has increasingly shifted toward large language model (LLM)-driven pipelines.
General-purpose research assistants such as Elicit, Consensus, and Scite facilitate abstract-level synthesis and citation-centric analysis, while tools like ChatPDF~\cite{ibrahim2024chatpdf} support document-level interaction.
However, these systems are not designed for large-scale, cross-corpus dataset discovery and lack explicit mechanisms for structured dataset indexing.

Recent work has begun to explore automated extraction of dataset references from scientific literature.
ChatPD~\cite{xu2025chatpd} introduces a three-stage pipeline, including paper collection, information extraction, and entity resolution, to construct structured paper–dataset networks, achieving high precision and mapping informal descriptions to standardized metadata schemas.
Similarly, Data Gatherer~\cite{marini-etal-2025-data} and related LLM-based IE systems~\cite{cheung2025llm} demonstrate that LLMs can reliably extract dataset mentions and metadata fields at scale, supporting FAIR (Findable, Accessible, Interoperable, and Reusable) principles.
Our work builds on these advances but differs in scope and objective. Rather than focusing on dataset extraction in isolation, we target urban data discovery as a first-class research problem. By systematically mining a decade of Nature-affiliated publications and organizing extracted datasets using an urban-specific metadata schema, we provide the first large-scale, literature-derived infrastructure for interdisciplinary urban data discovery.

\section{Paper2Data}\label{sec:pipeline}

This section presents our pipeline for large-scale urban dataset discovery from scientific literature.
As shown in Figure~\ref{fig:fram},  \textit{Paper2Data} consists of six steps:  \textit{Literature Curation}, 
\textit{Schema-guided Metadata Extraction}, 
\textit{Evidence-aware Data Verification}, 
\textit{Metadata Refinement and Harmonization}, 
\textit{External Resource Linking}, 
and \textit{Data Portal Construction}.
Each step is designed to be fully automated and scalable, enabling the continuous expansion of a unified urban data index.

\subsection{Literature Curation}\label{Sec:3.1}

To ensure information quality, we constructed a large-scale corpus of scientific publications from \textit{Nature}-affiliated journals, leveraging their broad disciplinary coverage and consistent article structure. The corpus spans a ten-year period from 2016 to 2025 and includes twelve general-interest journals (e.g., \textit{Nature}, \textit{Nature Cities}) and one data-centric journal (i.e., \textit{Scientific Data}). 

An automated HTML acquisition framework is designed to systematically collect publications from these venues (i.e., journal's official website).
As shown in Figure~\ref{fig:fram}~(step 1),  The framework first scans journal publication pages to retrieve the title and abstract of each article. An light LLM (i.e., Qwen3-4B) is then prompted as a detector to determine whether a publication is related to urban studies or involves the use of urban datasets.
For articles that satisfy this criterion, the full-text HTML is downloaded and parsed using structured HTML extraction, preserving all major components of the article, including the abstract, main body text, tables, figure captions, and supplementary descriptions.
The parsed content is subsequently converted into a structured text format for downstream metadata extraction.
In total, 15,726 full-text articles containing at least one urban dataset were collected. The number of publications from each journal is summarized in Table~\ref{tab:journals}. The broad coverage across topics ensures the diversity and richness of the resulting dataset.

\begin{table}[t]
\centering
\small
\caption{Nature-affiliated journals included in the corpus and the number of articles included for data extraction.}
\label{tab:journals}
\vspace{-2mm}
\begin{tabular}{lr}
\toprule
\textbf{Journal} & \textbf{\#} \\
\midrule
Nature & 245 \\
Nature Cities & 129 \\
Nature Climate Change & 258 \\
Nature Communications & 1,494 \\
Nature Computational Science & 19 \\
Nature Ecology \& Evolution & 16 \\
Nature Human Behaviour & 321 \\
Nature Machine Intelligence & 12 \\
Nature Physics & 4 \\
Nature Sustainability & 255 \\
npj Urban Sustainability & 199 \\
Humanities and Social Sciences Communications & 2,801 \\
Scientific Data & 19 \\
Scientific Reports & 9,954 \\\midrule
Totally    & 15,726 \\
\bottomrule
\end{tabular}
\vspace{-2mm}
\end{table}



\subsection{Schema-guided Metadata Extraction}\label{Sec:3.2}


To address the lack of a unified metadata schema and classification system for urban data, we design a schema-guided metadata extraction framework that systematically transforms unstructured dataset mentions in scientific articles into structured, machine-readable records suitable for large-scale discovery and indexing.

\paragraph{Metadata Schema Definition.}
We represent each identified dataset as a structured data card following a predefined schema. 
The schema is designed to capture the essential attributes required for dataset-level discovery, comparison, and reuse across disciplines, while remaining sufficiently general to accommodate heterogeneous urban data sources.
Formally, given an article $p$ with full text $T_p$, the extraction task aims to identify a set of dataset mentions $\{d_i\}$ and map each mention to a structured tuple:
\[
d_i = (\text{Name}, \text{Sum}, \text{Geo}, \text{Time}, \text{Type}, \text{URL}, \text{Ref}),
\]
presenting the dataset name, a concise summary, geographic and temporal coverage, category and sub-category information, access URLs for data download, or a specific reference to the publication introducing the dataset.


\paragraph{A Taxonomy of Urban Data Types.}

\begin{figure}[t]
  \centering
  \includegraphics[width=0.8\columnwidth]{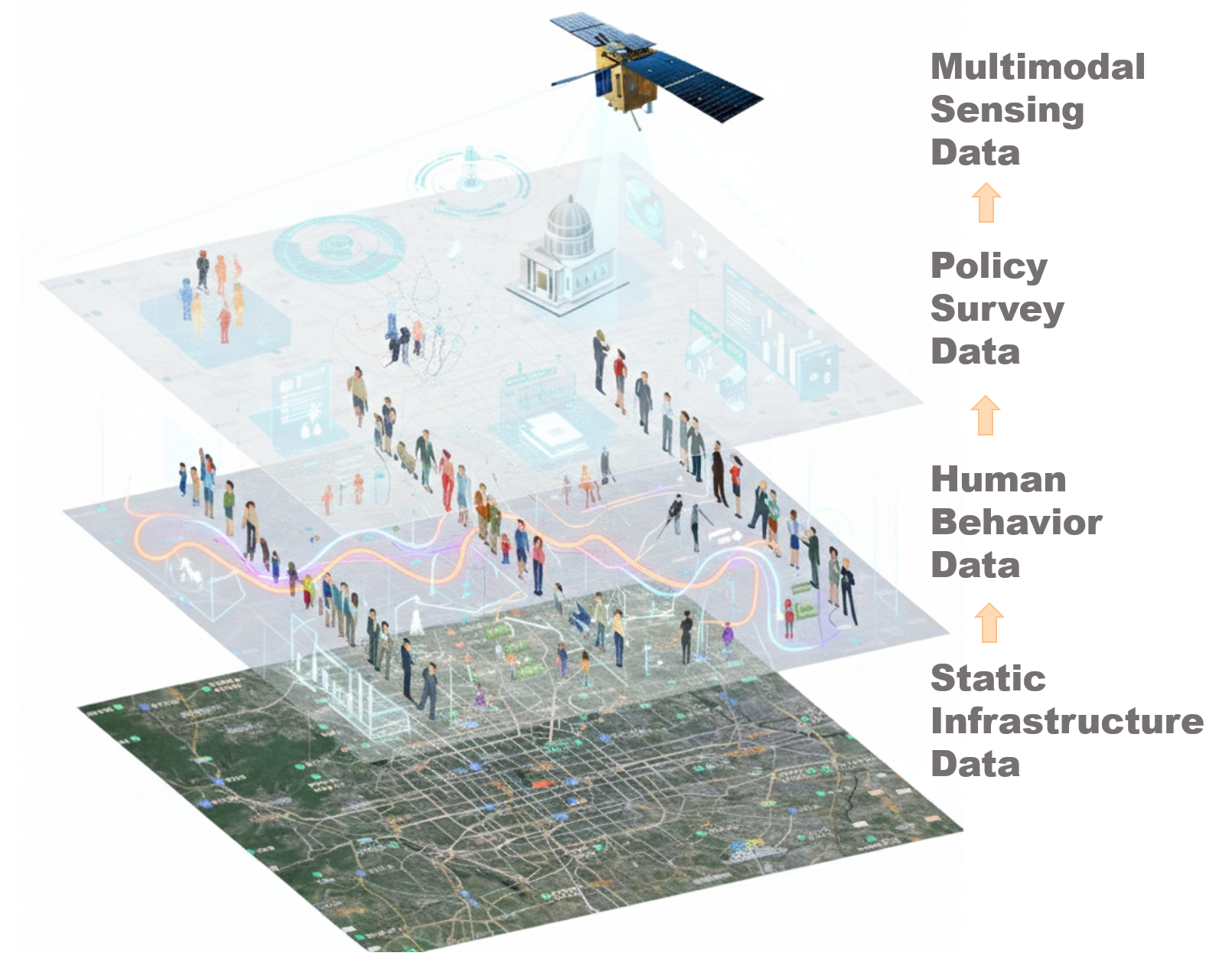}
  \caption{Urban data typology used for dataset indexing.}
  \label{fig:data_topylogy}
  \vspace{-3mm}
\end{figure}

While the metadata schema provides a standardized representation for individual datasets, effective discovery further requires a principled way to organize datasets across domains.
To this end, we introduce a four-category urban data taxonomy, illustrated in Figure~\ref{fig:data_topylogy}, which supports structured indexing and retrieval.
The taxonomy reflects fundamental dimensions of urban data, spanning from static to dynamic phenomena and from physical infrastructure to human activities, thereby capturing the layered structure of urban systems.

\begin{enumerate}
    \item \textbf{Statistical infrastructure data}: This group contains \textit{five} subcategories: \textit{road networks and transportation infrastructure}; \textit{building footprints and land-use maps}; \textit{points of interest (POIs)}; \textit{administrative boundaries and zoning maps}; and \textit{utility networks (electricity, water, and communication)}.
    These datasets describe relatively static or slowly evolving components of the urban built environment and serve as the structural backbone for spatial analysis and modeling.

    \item \textbf{Human behavior data}: This group contains \textit{four} subcategories: \textit{human mobility traces (GPS, transit cards, ride-hailing)}; \textit{socioeconomic activities (consumption, employment, and commerce)}; \textit{social media interactions and online behavior}; and \textit{health and wellbeing data (hospitalization counts and survey-based measures)}.
    This category captures dynamic, population-level processes that reflect how individuals and groups interact with the urban environment.

    \item \textbf{Policy and survey data}: This group contains \textit{four} subcategories: \textit{population censuses and household surveys}; \textit{statistical yearbooks and socioeconomic indicators}; \textit{government reports and urban planning documents}; and \textit{policy texts and regulatory frameworks}.
    These datasets encode institutional context, governance structures, and population-level attributes, providing essential background for causal inference and policy evaluation.

    \item \textbf{Multimodal sensing data}: This group contains \textit{five} subcategories: \textit{satellite remote sensing imagery (optical, SAR, night-time lights)}; \textit{aerial and drone imagery}; \textit{ground-based sensors (air quality, temperature, and noise)}; \textit{urban IoT devices (traffic, energy, water, environmental monitoring)}; and \textit{City-wide camera networks and meteorological stations}.
    This category captures high-resolution, often continuous measurements of the physical and environmental state of cities across space and time.
\end{enumerate}

Together, this taxonomy enables dataset retrieval based on shared mechanisms, variables, and spatial-temporal characteristics, rather than relying solely on discipline-specific or task-specific labels.

\paragraph{Extraction Method.}

Building on the metadata schema and taxonomy defined above, we employ an LLM to perform context-aware metadata extraction at the article level.
For each full-text article prepared as described in Section~\ref{Sec:3.2}, the LLM (Gemini-2.5-flash as discussed in Section~\ref{sec:results1}) is prompted to identify both explicit and implicit dataset mentions, distinguish original datasets from derived indicators or analytical methods, and normalize free-text descriptions into the predefined schema fields.

Because extraction is performed at the paper level, multiple datasets may be identified from a single article.
Each extracted dataset is returned as an individual data card in JSON format, and data cards generated from different articles can be merged seamlessly during parallel extraction. To facilitate downstream metadata verification, we also require the LLM to provide \textit{evidence of each dataset}. Specifically, the model returns verbatim text spans from the original article that describe the specific data, together with their section locations.
This design enables systematic validation of extracted metadata against the source document. For clarity, we present a simplified version of the extraction prompt an example metadata (e.g., from~\cite{Althoff2025countrywide}) below.

\begin{PromptBox}{Prompt for Metadata Extraction (Simplified)}
You are a research assistant tasked with extracting dataset metadata from a full academic article.
Given the complete article text, identify all  **original datasets** used in the study.
For each dataset, generate a structured record following the **predefined schema**, including its name, category, temporal and geographic coverage, access information, and a concise description of the data itself.

If multiple datasets are used, return one record per dataset. Output only a valid **JSON** array, with each dataset represented as a JSON object. For each field, provide a short verbatim text excerpt from the article as **evidence** to support verification.
\end{PromptBox}
\begin{JsonBox}{Extracted Metadata (an Example)}
"paper_title": "Countrywide natural experiment links built environment to physical activity",
"meta_data": [{
    "dataset_id": "ds-01",
    "Data_Name": "Walk Score walkability scores",
    "Data_summary": "City-level walkability scores used to characterize the built environment of origin and destination cities in relocation analyses. The paper uses Walk Score values (scale 1 to 100) for US cities; the scores are described as based on nearby amenities within a 0.25-mile to 1.5-mile radius (with distance decay) and pedestrian-friendliness measures such as block length and intersection density, and are used to compute walkability differences between cities.",
    "Category": "Statical Infrastructure Data",
    "Sub-category": "Road networks and transportation infrastructure",
    "Time_Coverage": "March 2013 to February 2016",
    "Geographic_Coverage": "USA (city-level scores for the 1,609 cities in the study)",
    "URL": "https://github.com/behavioral-data/movers-public",
    "ref": ["Cities & Neighborhoods. Walk Score www.walkscore.com/cities-and-neighborhoods/ (accessed 17 June 2018)."],
    "Other_Information": "Evidence: Scores are on a scale of 1 to 100 where 100 is the most walkable.; Location: Methods, Walkability measure; Confidence: high"},]
\end{JsonBox}

\vspace{-10mm}
\subsection{Evidence-aware Data Verification}\label{Sec:3.3}

A key challenge in applying large language models to long scientific articles is hallucination, where plausible but unsupported dataset attributes may be generated.
To mitigate this issue, we introduce an evidence-aware data verification step that explicitly grounds extracted metadata in the original article text.
During metadata extraction, the LLM is required to return verbatim text spans as evidence for each metadata field.
We first \textit{locate these evidence spans in the source document} using rule-based exact and fuzzy string matching, ensuring that each extracted attribute is anchored to a concrete textual context.
We then \textit{perform semantic verification} by comparing the extracted metadata with the localized source text to assess whether the claimed attribute is genuinely supported by the article.
Metadata fields that fail evidence localization or semantic consistency checks are filtered out.
This verification step substantially reduces hallucinated metadata and improves the factual reliability of the extracted dataset records.

\subsection{Metadata Refinement and Harmonization}\label{Sec:3.4}

Automatically extracted metadata may contain inconsistencies, missing fields, or non-standard representations.
To improve reliability and support downstream indexing, we introduce a metadata refinement and harmonization module that standardizes and validates extracted records before portal construction.
Specifically, this module performs:
(i) temporal normalization by converting free-text time descriptions into standardized year or year--month formats;
(ii) geographic normalization by mapping location mentions to canonical country and administrative region levels;
(iii) category validation to ensure that each dataset is assigned a valid and non-missing data type within the urban data taxonomy; and
(iv) reference checking to verify that dataset citations or source publications are valid and correctly linked.
These steps combine rule-based normalization with lightweight LLM-assisted cross-validation using both the dataset summary and the original article text, filtering malformed or weakly supported records and reducing false positives prior to indexing.
This step ensures that the final corpus supports consistent filtering, aggregation, and cross-dataset comparison.

\subsection{External Resource Linking}\label{Sec:3.5}

During dataset extraction, we observe that URLs mentioned in scientific articles are frequently \textit{outdated} or \textit{no longer accessible} due to link rot or repository reorganization.
To improve data accessibility without introducing incorrect links, we introduce an external resource linking step implemented as a lightweight LLM-driven agent.
Based on the structured metadata schema, the agent first constructs a concise dataset description using the extracted name, category, geographic and temporal coverage, and source context.
This description is then used to query a web search API to retrieve candidate access links.
The agent evaluates these candidates by jointly considering semantic relevance to the dataset description and consistency with the original paper context, and selects a replacement URL only when sufficient confidence is achieved.
When no reliable alternative is identified, the original literature reference is retained as a fallback.
This conservative, agent-mediated linking strategy improves the practical usability of dataset records while minimizing the risk of introducing spurious external resources.

\subsection{Urban Data Discovery Portal}\label{Sec:3.6}

All verified and harmonized data cards are indexed and served through an open urban data discovery portal, \textit{UrbanDataMiner}.
The portal treats datasets as first-class objects and supports dataset-level search and filtering across multiple dimensions.
Specifically, users can perform keyword-based search over dataset names and summaries, apply category-based filtering using the urban data taxonomy, and refine results by geographic coverage (country or administrative region) and temporal coverage.
These structured filters enable efficient, hypothesis-driven dataset discovery without requiring manual inspection of individual papers.

To support exploratory and semantically driven queries, we further implement a retrieval-augmented generation (RAG) interface.
User queries are embedded and matched against dataset summaries in the unified corpus, allowing relevant datasets to be retrieved even when exact keywords are not explicitly specified.
The RAG interface returns grounded dataset records rather than free-form text, ensuring that all retrieved results remain traceable to structured metadata and source literature.

\section{Experiments}\label{sec:results}

\subsection{Human-Annotated Benchmark Evaluation}

To validate the effectiveness of our pipeline, we address the following two research questions:
\begin{itemize}
    \item \textbf{RQ1:} To what extent can our pipeline identify research datasets used in the literature?
    \item \textbf{RQ2:} To what extent does the verified metadata align with human annotations at the field level?
\end{itemize}
\vspace{-3mm}
\subsubsection{Benchmark.}\label{sec:results1}
To answer these questions, we construct a human-annotated benchmark.
Specifically, three to five papers are randomly sampled from each journal listed in Table~\ref{tab:journals}, yielding a total of \textbf{54 papers}.
Two domain experts independently review each paper and annotate all research datasets in strict accordance with the schema defined in Section~\ref{Sec:3.2}.
The two annotation sets are then compared, and any disagreements are resolved through face-to-face discussion to reach a consensus, ensuring high annotation fidelity.
In addition, each identified dataset is assigned a two-level relevance label:
\textbf{L1} denotes datasets that serve as core resources for the main contribution of the paper;
\textbf{L2} denotes datasets that are related to the study but not central to the primary analysis;
datasets with only peripheral mentions are not explicitly labeled.
In total, the benchmark contains \textbf{307 datasets} annotated with L1 or L2 relevance.

\vspace{-1mm}
\subsubsection{Evaluation Protocol.}
To address the two research questions, we design a three-stage matching protocol to align pipeline-extracted datasets with benchmark annotations.
For each benchmark dataset within a paper, we first retrieve a set of candidate datasets extracted by our pipeline.
Specifically, a sentence embedding model~\cite{chen2024bge_m3_embedding} is used to compute semantic similarity between the benchmark dataset summary and all extracted dataset summaries from the same paper, and the top-5 most similar candidates are selected.
Next, we employ an LLM-as-judge (GPT-5 as the base model) to perform a fine-grained comparison between the benchmark dataset summary and each candidate extracted summary, determining whether they refer to the same underlying dataset.
If a match is identified, the LLM-as-judge is further applied to assess field-level consistency between the extracted metadata and the benchmark annotations.

We note that datasets with broad geographic or temporal coverage may be described either as a single unified dataset or as multiple related subsets.
To account for this ambiguity, we report results under both \textit{expanded} and \textit{strict} matching protocols.
Under \textit{strict}, each extracted dataset must exclusively and closely match a single benchmark dataset.
Under \textit{expanded}, multiple extracted records are allowed to jointly match one benchmark dataset, capturing subset or containment relationships.
For \textbf{RQ1}, we report recall evaluated on both L1 datasets and the full benchmark dataset set (L1+L2). Different LLM backbones in our pipeline are compared. 
For \textbf{RQ2}, based on the best LLM backbone,  we report field-level accuracy computed over its successfully matched dataset pairs.

\subsubsection{Results.}
The results for \textbf{RQ1} are summarized in Table~\ref{tab:main_results}.
Across different LLM backbones, we observe clear performance variation, suggesting that metadata extraction from full-length scientific papers is a challenging task.
Lightweight models such as GPT-4o-mini exhibit limited effectiveness in this setting.
Among the evaluated models, GPT-5 yields the highest recall, while Gemini-2.5-flash achieves similar recall with significantly lower time and cost. Based on the efficiency-performance trade-off, we select \textbf{Gemini-2.5-flash} as the representative backbone for answering this question.
Using Gemini-2.5-flash, the pipeline achieves a recall of \textbf{92.64\%} for core datasets (L1) and \textbf{88.93\%} for all annotated datasets (L1+L2) under the expanded matching protocol.
Even under the stricter matching criterion, recall remains above \textbf{67\%} for L1 datasets.
These results indicate the majority of research datasets explicitly or implicitly used in the literature can be successfully identified by the pipeline, including datasets that are not central to the main contribution but are nonetheless relevant for reuse.

\begin{table}[t]
\centering
\small
\caption{Data identification recall (\%) under different matching protocols and cost measurements.}
\label{tab:main_results}
\vspace{-2mm}
\setlength{\tabcolsep}{4pt}
\begin{tabular}{lcccccc}
\toprule
\multirow{2}{*}{\textbf{Model}} &
\multicolumn{2}{c}{\textbf{L1}} &
\multicolumn{2}{c}{\textbf{L1+L2}} &
\multirow{2}{*}{\textbf{Time}} &
\multirow{2}{*}{\textbf{Cost}} \\
\cmidrule(lr){2-3}\cmidrule(lr){4-5}
& \textbf{Strict} & \textbf{Exp.} & \textbf{Strict} & \textbf{Exp.} & & \\
\midrule
Gemini-2.5-flash & 67.48 & \cellcolor{pink}92.64 & 63.84 & \cellcolor{pink}88.93 & 2m49s & \cellcolor{pink}\$0.64 \\
Gemini-3-flash   & 64.42 & 86.50 & 60.91 & 85.99 & 36m47s & \$5.82 \\
DeepSeek-V3      & 68.66 & 86.57 & 57.01 & 74.77 & 3m54s & \$0.16 \\
DeepSeek-V3.2    & 71.78 & 89.57 & 65.47 & 86.97 & 10m34s & \$0.35 \\
GPT-4o-mini      & 64.42 & 81.60 & 49.51 & 68.08 & 1m23s & \$0.15 \\
GPT-5            & 74.23 & \cellcolor{gray!15}94.48 & 71.34 & \cellcolor{gray!15}91.86 & 4m54s & \cellcolor{gray!15}\$3.50 \\
\bottomrule
\end{tabular}
\vspace{-0mm}
\end{table}

\begin{table}[t]
\centering
\small
\caption{Field-level accuracy (\%) before and after metadata refinement. Full results can be found in Appendix Table~\ref{tab:field_accuracy_models_2}.}
\label{tab:field_accuracy_models}
\vspace{-2mm}
\setlength{\tabcolsep}{4pt}
\begin{tabular}{lllll}
\toprule
\multirow{2}{*}{\textbf{Field}} &
\multicolumn{2}{c}{\textbf{Before Refinement}} &
\multicolumn{2}{c}{\textbf{After Refinement}} \\
\cmidrule(lr){2-3}\cmidrule(lr){4-5}
& \textbf{Strict} & \textbf{Expanded} & \textbf{Strict} & \textbf{Expanded} \\
\midrule
Time Coverage        & 55.61 & 71.79 & 66.33\textcolor{red}{\(\uparrow\)} & 90.84\textcolor{red}{\(\uparrow\)} \\
Geographic Coverage  & 78.57 & 98.17 & 80.10 & 98.90 \\
Category              & 80.61 & 79.85 & 84.69\textcolor{red}{\(\uparrow\)} & 82.78\textcolor{red}{\(\uparrow\)} \\
Sub-category          & 67.34 & 65.57 & 70.91\textcolor{red}{\(\uparrow\)} & 70.33\textcolor{red}{\(\uparrow\)} \\
URL                   & 85.71 & 85.35 & 85.71 & 85.35 \\
Reference             & 83.67 & 86.81 & 82.65 & 89.01\textcolor{red}{\(\uparrow\)} \\
\bottomrule
\end{tabular}
\vspace{-3mm}
\end{table}

The results for \textbf{RQ2} are reported in Table~\ref{tab:field_accuracy_models}, comparing field-level accuracy before and after metadata refinement.
We observe that the initial extraction already achieves relatively high accuracy for explicitly stated fields such as \textit{geographic Coverage}, \textit{URL}, and \textit{reference}, with limited changes after refinement.
In contrast, fields that are more ambiguously described in free text, including \textit{Time Coverage} and categorical attributes, exhibit substantially lower accuracy prior to refinement but improve consistently after the refinement step.
For example, the expanded accuracy of \textit{time Coverage} increases from 71.79\% to 90.84\%.
Overall, the final accuracy across the six key metadata fields ranges from 70\% to 90\% (\textbf{86.2\% in average}), indicating that the proposed refinement module effectively improves metadata precision for downstream dataset discovery and reuse.

\subsection{Comparison with Search Engines}

To demonstrate the advantage of our system in urban dataset discovery, we compare the search performance of the \textit{UrbanDataMiner} portal with six widely used general-purpose search engines.
For each target dataset in the benchmark, we construct a standardized query by combining the dataset name with its geographic and temporal coverage, thereby specifying a concrete and well-scoped data requirement.
The same query is issued to all systems, and the returned results are subsequently evaluated through semantic matching against the benchmark dataset descriptions.

Based on this evaluation, we report three metrics:
(1) \textit{\#Match}, the number of datasets whose returned top 10 results (including papers) can be semantically matched to the target benchmark dataset;
(2) \textit{\#URL}, the number of datasets for which a correct and usable dataset access URL is retrieved, defined as a URL whose associated description is semantically consistent with the benchmark dataset summary; and
(3) \textit{Avg. Rank}, the average rank position of the correct dataset URL among the top-10 returned results.
The comparative results are summarized in Table~\ref{tab:union_summary}.

\begin{table}[t]
\centering
\caption{Comparison of dataset discovery performance between UrbanDataMiner and general-purpose search APIs.}
\label{tab:union_summary}
\vspace{-2mm}
\small
\begin{tabular}{lcccc}
\toprule
\textbf{System} & \textbf{\#Match (\%)} & \textbf{\#URL (\%)} & \textbf{Avg. rank}  \\ \midrule
Google & 190 (61.89) & 164 (53.42) & 2.23  \\
Bing & 205 (66.78) & 178 (57.98) & 2.56 \\
Brave & 182 (59.28) & 161 (52.44) & 2.52 \\
DuckDuckGo & 210 (68.40) & 180 (58.63) & 2.98  \\
Yahoo & 201 (65.47) & 179 (58.31) & 2.13  \\
Tavily & 201 (65.47) & 171 (55.70) & 3.78  \\
\midrule
Union & 243 (79.15) & 204 (66.45) & 1.40  \\
\midrule
benchmark & 307 (100.00) & 199 (64.82) & -  \\
benchmark (valid) & 307 (100.00) & 187 (60.91)  & 1.00\\
UrbanDataMiner (Raw) & 271 (88.27) & 173 (56.35) & -  \\
UrbanDataMiner (Valid) & 271 (88.27) & 110 (35.83) & 1.34  \\
UrbanDataMiner (Final) & 271 (88.27) & 221 (71.99 ) & 1.30  \\
\bottomrule
\end{tabular}
\vspace{-5mm}
\end{table}





\subsubsection{Overall Results.}
Table~\ref{tab:union_summary} demonstrates a clear advantage of \textit{UrbanDataMiner} in comprehensive dataset discovery.
While individual general-purpose search engines identify approximately 60\% of the benchmark datasets, and their union achieves 79\% coverage, \textit{UrbanDataMiner} attains complete coverage by successfully identifying 88.27\% of the benchmark datasets.
This result indicates that \textit{UrbanDataMiner} can additionally retrieve \textbf{around 9\% of datasets} that are not easily discoverable through general-purpose search engines.

Regarding dataset access, although general-purpose search engines may return a larger number of links, these links often require substantial manual effort to locate the correct dataset.
By contrast, \textit{Paper2Data}, together with the external resource linking step, effectively leverages these resources to recover valid dataset URLs, enabling more reliable dataset identification and easier access in practice.


\subsubsection{Additional Analysis.}
We further analyze cases where \textit{UrbanDataMiner} successfully retrieves a target dataset while general-purpose search engines and APIs fail. We identify two primary reasons for this advantage. 
First, \textit{UrbanDataMiner} is able to \textbf{directly surface dataset-specific URLs} from large data platforms when these datasets are explicitly introduced in the scientific literature. For example, for the query \emph{``Global PRISMA hyperspectral satellite data from 2019 to 2024''}, Google returns only the general PRISMA mission webpage (\url{https://www.asi.it/en/earth-science/prisma/}). In contrast, \textit{UrbanDataMiner} retrieves a precise dataset landing page with a persistent identifier (\url{https://doi.org/10.6084/m9.figshare.29247518.v2}), as the dataset is explicitly described and cited in the corresponding paper~\cite{Zhou2025}.
Second, \textit{UrbanDataMiner} can \textbf{identify datasets that are introduced only implicitly or privately within papers} and are therefore not easily discoverable through keyword-based web search.
For example, when querying ``Pre-intervention gunshot victimization and arrest covariates (2007--2018),''
\textit{UrbanDataMiner} returns the specific reference~\cite{SteinmanZimmerman2003} and the contact information~\cite{WoodPapachristos2019} for requesting access to the data.
In contrast, Google returns only a small number of loosely related resources, from which identifying the exact dataset still requires substantial manual exploration.
These cases highlight the benefit of literature-grounded dataset extraction: by leveraging paper-level context and explicit data mentions, \textit{UrbanDataMiner} enables more precise and actionable dataset discovery than generic search engines.

\section{Analysis of Discovered Urban Datasets}\label{sec:data}

\begin{figure*}[t]
  \centering
  \includegraphics[width=\textwidth]{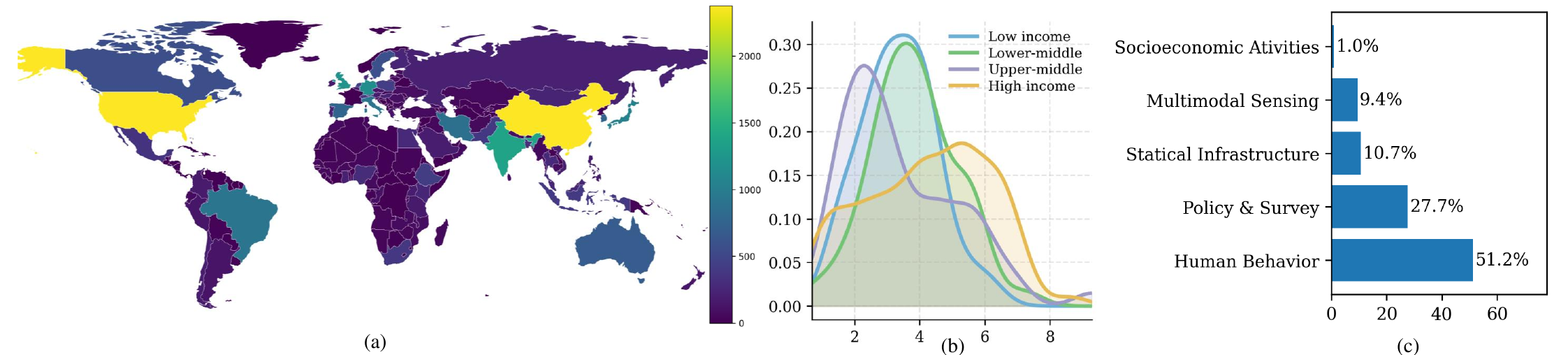}\par
  \vspace{0pt} 
  \includegraphics[width=\textwidth]{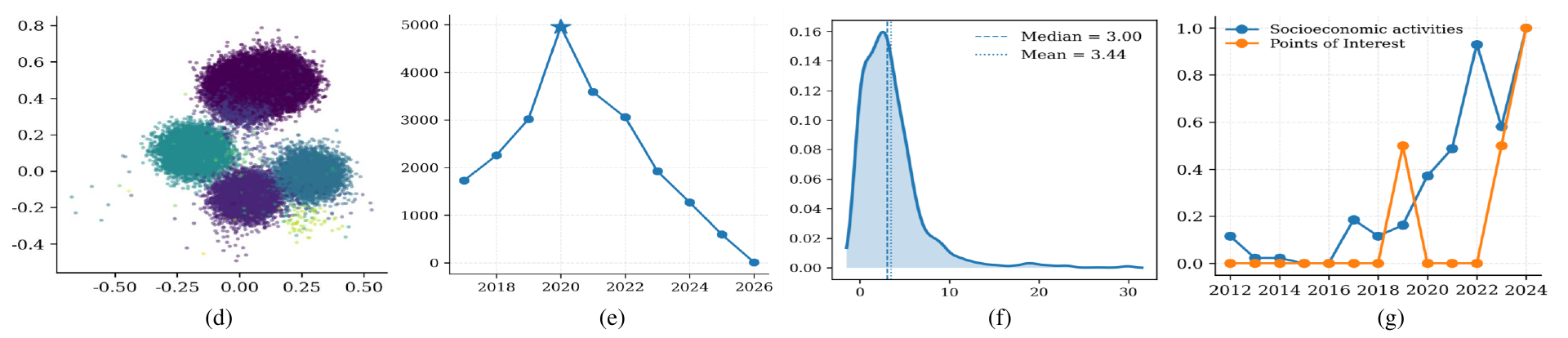}
\vspace{-5mm}
  \caption{Distributional analysis of urban data.
(a) Global geographic concentration of data across regions.
(b) Distributional disparities across country income groups.
(c) Category composition and prevalence.
(d) Semantic embedding and clustering of data categories, revealing a core--periphery structure of major urban data domains.
(e) Temporal distribution of urban datasets by data collection year.
(f) Distribution of publication latency between data collection and publication.
(g) Fastest-growing urban data domains after adjusting for publication delay.
}
  \label{fig:global_analysis}
\vspace{-2mm}
\end{figure*}

In this section, we conduct an in-depth analysis of the 65{,}632 urban datasets discovered by our system, providing insights into how urban data are used in research, as well as their geographic distribution across countries and temporal trends.

\subsection{Spatial Distribution}
As shown in Figure~\ref{fig:global_analysis}(a), the data are highly concentrated in a small number of countries, with China and the United States serving as the dominant sources. A second tier includes India, several countries in Western Europe, and other high-income economies such as Australia. We further observe a correlation between national scale and data volume, where countries with larger populations tend to contribute more data. 

To further examine structural disparities beyond national scale, we analyze the distribution across income groups. As shown in Figure~\ref{fig:global_analysis}(b), the distribution reveals a clear stratification effect, with upper-middle-income countries exhibiting the lowest overall data availability. In other words, data availability does not increase monotonically with income level. One possible explanation lies in structural differences in data production across development stages. Large-scale international survey programmes, such as the Demographic and Health Surveys (DHS) covering dozens of low- and middle-income countries, provide standardized and publicly accessible data for many low-income settings \citep{nature_dhs}. In contrast, high-income countries benefit from mature statistical systems, well-established research infrastructures, and sustained data collection mechanisms \citep{nature_globalnorth}. Upper-middle-income countries may fall between these two regimes: they often receive fewer externally funded data initiatives, while their domestic statistical and research systems may not yet reach the maturity of high-income economies, resulting in a structural “missing middle” pattern in data availability.

\subsection{Category Distribution}
As shown in Figure~\ref{fig:global_analysis}(c), we present the five most frequent data categories. Human behavior data constitute the largest portion, accounting for over half of the records, followed by policy and survey data as the second most prevalent category. Together, these two types make up nearly 80\% of the dataset. Infrastructure-related and multimodal sensing data jointly account for roughly 20\%, while socioeconomic activity data remain relatively scarce.
As shown in Figure~\ref{fig:global_analysis}(d), we first embed category labels into a semantic space and then perform clustering to identify distinct data domains. The clustering results reveal a two-level semantic structure of the data taxonomy. The core semantic region consists of four large clusters, corresponding to the major domains discussed earlier: human behavior, social surveys, infrastructure, and multimodal sensing.

In contrast, the peripheral semantic region is not composed of random or noisy categories; instead, it is consistently dominated by tooling-oriented concepts and privacy-sensitive or regulated domains. The most scarce data domains include cryptocurrency-related data and clinical medical data. Due to their high privacy sensitivity and regulatory constraints, these domains are less likely to form large-scale, reusable datasets.

\subsection{Temporal Trends} 
As shown in Figure~\ref{fig:global_analysis}(e), we visualize the temporal distribution of urban datasets. The data volume reaches a pronounced peak around 2020, which can be attributed to the massive surge of data generated during the COVID-19 crisis. This surge spans multiple domains, including government policy data~\citep{nature_covid_govern}, large-scale social and psychological surveys~\citep{nature_covid_psychological}, and individual-level public health records~\citep{nature_covid_3}.

We further observe a steady decline in the data volume from 2020 to 2025. However, we argue that this trend reflects a statistical latency effect rather than a genuine reduction in data production. As illustrated in Figure~\ref{fig:global_analysis}(f), the average gap between the data collection year and the publication year is approximately 3.44 years. Therefore, the apparent post-2020 decline should not be interpreted as a true structural contraction of data generation, but rather as an artifact of delayed data release. Furthermore, as shown in Figure~\ref{fig:global_analysis}(g), we analyze the fastest-growing domains of urban data. Given the significant publication latency inherent in many urban datasets, we restrict the analysis to data collected up to 2024 to reduce statistical noise caused by delayed releases.

The results show that both socioeconomic activity data and points-of-interest (POI) data exhibit sustained growth. Driven by the expansion of mobile internet services and the development of digital cities, the structure of urban data is gradually shifting away from traditional human behavior and survey-based data toward more fine-grained representations of economic activities and spatial semantics.

\vspace{-2mm}
\section{Conclusion}

In this paper, we present \textit{Paper2Data} and \textit{UrbanDataMiner}, a large-scale, literature-derived resource for urban dataset discovery.
By extracting dataset mentions and structured metadata from scientific publications, our work addresses a key limitation of document-centric search and enables dataset-level retrieval at scale.
Applied to over 15,000 Nature-affiliated articles, the resulting corpus comprises more than 65,632 urban dataset records, supported by a unified metadata schema, human-annotated evaluation, and open access through an online portal.
We believe this resource will facilitate dataset discovery, reuse, and meta-analysis across urban science and related domains, and provide a foundation for future research on large-scale data-centric knowledge infrastructures.


\bibliographystyle{ACM-Reference-Format}
\bibliography{sample-base}


\appendix

\section{Additional Results}
\begin{table*}[t]
\centering
\small
\caption{Field-level accuracy (\%) with counts in parentheses.}
\label{tab:field_accuracy_models_2}
\vspace{-2mm}
\resizebox{0.9\textwidth}{!}{%
\begin{tabular}{lcccc}
\toprule
\multirow{2}{*}{\textbf{Field}} &
\multicolumn{2}{c}{\textbf{Before Refinement}} &
\multicolumn{2}{c}{\textbf{After Refinement}} \\
\cmidrule(lr){2-3}\cmidrule(lr){4-5}
& \textbf{Strict  Recall for L1+L2} & \textbf{Expanded  Recall for L1+L2} & \textbf{Strict  Recall for L1+L2} & \textbf{Expanded  Recall for L1+L2} \\
\midrule
Time\_Coverage        & 55.61 (109/196) & 71.79 (196/273) & 66.33 (130/196) &  90.84 (248/273) \\
Geographic\_Coverage  & 78.57 (154/196) & 98.17 (268/273) & 80.10 (157/196) & 98.90 (270/273) \\
Category              & 80.61 (158/196) & 79.85 (218/273) & 84.69 (166/196) & 82.78 (226/273)\\
Sub-category          & 67.34 (132/196) & 65.57 (179/273) & 70.91 (139/196) & 70.33 (192/273) \\
URL                   & 85.71 (168/196)&  85.35 (233/273) & 85.71 (168/196)&  85.35 (233/273) \\
ref                   &  83.67 (164/196) & 86.81 (237/273) & 82.65 (162/196) & 89.01 (243/273) \\
\bottomrule
\end{tabular}%
}
\end{table*}

\end{document}